\documentclass[11pt,a4paper]{article}
\usepackage{jheppub}
\usepackage{graphicx,amsfonts,amssymb,amsmath}

\begin{document}

\title{Holographic View on Quantum Correlations and Mutual Information between Disjoint Blocks of a Quantum Critical System.}
\author[a]{Javier Molina-Vilaplana,}
\emailAdd{javi.molina@upct.es}
\affiliation[a]{Department of Systems Engineering and Automation. Technical University of Cartagena \\ C/ Dr Fleming SN 30202, Cartagena, Spain}

\author[b,1]{Pasquale Sodano \note{Permanent Address: Dipartimento di Fisica, Universit\'a di Perugia, Via A. Pascoli, I-06123, Perugia, Italy}}
\affiliation[b]{Perimeter Institute of Theoretical Physics, 31 Caroline Street North,
Waterloo, ON, N2L2Y5, Canada.}

\abstract{In ($d+1$) dimensional \emph{Multiscale Entanglement Renormalization Ansatz} 
(MERA) networks, tensors are connected so as to reproduce the discrete, ($d+2$) holographic 
geometry of Anti de Sitter space (AdS$_{d+2}$) with the original system lying at the boundary. 
We analyze the MERA renormalization flow that arises when computing the quantum correlations between two disjoint blocks of a quantum critical system, to show that the structure of the causal cones characteristic of MERA, requires a transition between two different regimes attainable by changing the ratio between the size and the separation of the two disjoint blocks. We argue that this transition in the MERA causal developments of the blocks may be easily accounted by an AdS$_{d+2}$ black hole geometry when the mutual information is computed using the Ryu-Takayanagi formula. As an explicit example, we use a BTZ AdS$_3$ black hole to compute the MI and the quantum correlations between two disjoint intervals of a one dimensional boundary critical system. Our results for this low dimensional system not only show the existence of a phase transition emerging when the conformal four point ratio reaches a critical value but also provide an intuitive entropic argument accounting for the source of this instability. We discuss the robustness of this transition when finite temperature and finite size effects are taken into account.} 


\maketitle

\section{Introduction}\label{intro}
Entanglement entropy (EE) is by now regarded as a valuable tool to
witness the amount of entanglement in quantum field theories and
many body systems. By partitioning a given system ${\mathcal S}$
into two complementary sets ${\mathcal A}$ and
$\widetilde{\mathcal A}$ such that ${\mathcal S}= {\mathcal A}
\cup \widetilde{\mathcal A}$, the reduced density matrix
$\rho_{\mathcal A}$ (\textit{i.e}, the density matrix for an
observer accessing only the degrees of freedom of the subsystem
${\mathcal A}$), is obtained by tracing the full density matrix
$\rho$ over the degrees of freedom contained in
$\widetilde{\mathcal A}$ \textit{i.e}, $\rho_{\mathcal A} =
\rm{Tr}_{\widetilde{\mathcal A}} (\rho) $. The EE accounts for the
amount of quantum correlations between the complementary regions
${\mathcal A}$ and $\widetilde{\mathcal A}$ and is defined as the
von Neumann entropy of $\rho_{\mathcal A}$,

\begin{equation}
S_{\mathcal A}=-\rm{Tr}(\rho_{\mathcal A} \log \rho_{\mathcal A})~.
\label{eent}
\end{equation}

A standard approach to compute the entanglement entropy makes use
of the replica trick \cite{Holzhey94, cargliozzi, Calabrese04}.
The replica trick may be applied when the density matrix for the
full system is represented by a path integral (as in the vacuum or
in a thermal state); then, one can rather easily obtain the EE
(\ref{eent}) of the subsystem ${\mathcal A}$ from the knowledge
of,
\begin{equation}
S_{\mathcal A}=-\frac{\partial}{\partial n} {\rm Tr} \rho_{\mathcal A}^n \vert_{n=1}~.
\label{reptrick}
\end{equation}
In \cite{Calabrese04}, it has been shown that, for $d=1$ quantum
critical models, ${\rm Tr} \rho_{\mathcal A}^n = c_n
(\ell_{\mathcal A} / \epsilon)^{-(c/6)(n - 1/n)}$, where
$\ell_{\mathcal A}$ is the length of the interval ${\mathcal A}$,
$c$ is the central charge of the conformal field theory (CFT)
describing the given system at criticality and $\epsilon$ is an
ultraviolet cutoff. Using (\ref{reptrick}), one obtains that the
EE is given by,

\begin{equation}
S_{\mathcal A} = \frac{c}{3} \log \left( \frac{\ell_{\mathcal A}}{\epsilon}\right) + s_1~,
\label{eent_sint}
\end{equation}
where $s_1$ is a non universal constant.

Using an alternative approach based on holography Ryu and
Takayanagi (RT) derived a celebrated formula yielding the EE of
the region ${\mathcal A}$ provided that the (boundary) conformal
field theory describing the critical system admits an holographic
gravity dual \cite{Ryu106,Ryu06}. In the RT approach, the EE is
obtained from the computation of a minimal surface in the dual
higher dimensional gravitational geometry (bulk theory); as a
result, the entanglement entropy $S_{\mathcal A}$ in a CFT$_{d+1}$
is given by the celebrated area law relation,

\begin{equation}
S_{\mathcal A}=\frac{{\rm Area}(\gamma_{\mathcal A})}{4G^{(d+2)}_N}~,
\label{arealaw}
\end{equation}
where $d$ is the spatial dimension of the boundary CFT,
$\gamma_{\mathcal A}$ is the $d$-dimensional static minimal
surface in AdS$_{d+2}$ whose boundary and area are given by
$\partial {\mathcal A}$ and ${\rm Area}(\gamma_{\mathcal A})$,
respectively. $G^{(d+2)}_{N}$ is the $d+2$ dimensional Newton
constant. The RT proposal is physically appealing since looking
for the minimal surface $\gamma_{\mathcal A}$ separating the
degrees of freedom contained in region ${\mathcal A}$ from those
contained in $\widetilde{\mathcal A}$ amounts to search for the
severest entropy bound on the information hidden in the
AdS$_{d+2}$ region related with $\widetilde{\mathcal A}$. For
$d=1$, eq. (\ref{arealaw}) becomes \cite{ Ryu106},

\begin{equation}
S_{\mathcal A}=\frac{{\rm Length}(\gamma_{\mathcal A})}{4G^{(3)}_N}~.
\label{RTarealaw}
\end{equation}

Although the RT formula has not been rigorously proven its
validity is supported by very comforting evidence.\footnote{See \cite{Fursaev06, CasHuMyers2011} for some interesting attempts to derive it.} For instance,
one may show \cite{headrick07} that it provides a simple tool to
prove the strong subadditivity of EE, \emph{i.e} given two regions
$A$ and $B$,
\begin{equation}\label{strongsub}
S_{A} + S_{B}\geq S_{A \cup B} + S_{A \cap B}~;
\end{equation}
furthermore, eq. (\ref{arealaw}) together with (\ref{strongsub})
may be used also (at least in the context of strongly coupled
gauge theories, \textit{i.e} at a t'Hooft coupling $\lambda\gg 1$)
to derive the concavity property of coplanar Wilson loops defined
on curves $C_{A}=\partial A$ and $C_{B}=\partial B$ lying in the
same two dimensional plane. Namely,
\begin{equation}\label{wiloop}
\left\langle W(C_{A})\right\rangle \left\langle W(C_{B})\right\rangle \leq \left\langle W(C_{A \cup B})\right\rangle \left\langle W(C_{A \cap B})\right\rangle~,
\end{equation}
where $C_{A \cup B}=\partial(A \cup B)$ and $C_{A \cap
B}=\partial(A \cap B)$. To derive (\ref{wiloop}) one only needs to
note that, from the Maldacena conjecture \cite{AdSCFTbible00}, the
expectation value of a Wilson loop defined along a curve $C$ is
related to the area of the minimal surface $\gamma$ bounded by $C$
by,
\begin{equation}\label{holowiloop}
\left\langle W(C)\right\rangle \simeq \exp (-\sqrt{\lambda} \rm{Area}(\gamma))~.
\end{equation}
If in (\ref{holowiloop}) one takes $C=C_{A}=\partial A$ and
$\gamma=\gamma_{A}$, using (\ref{arealaw}) one can establish, up
to a constant that $S_A \sim -\log \left\langle W(\partial
A)\right\rangle$. Similar arguments yield $S_B \sim -\log
\left\langle W(\partial B)\right\rangle$, $S_{A \cup B} \sim -\log
\left\langle W(\partial (A \cup  B))\right\rangle$ and $S_{A \cap
B} \sim -\log \left\langle W(\partial (A \cap  B))\right\rangle$.
As a result, using (\ref{strongsub}), one gets (\ref{wiloop}).

The minimal curves used in the RT formula, allow also to compute
the two point functions of conformal primary operators of ${\rm
CFT}_{d+1}$ with an holographic gravity dual that is an
asymptotically ${\rm AdS}_{d+2}$ space-time. The holographic
computation of the correlation functions of these operators yields
\cite{susskind98},

\begin{equation}
\langle {\cal O}(x_{i}) {\cal O} (x_{j}) \rangle \sim \exp (-\Delta \, {\rm Length}(\gamma_{ij}))~,
\label{holocorr}
\end{equation}
where $\Delta$ is the operator scaling dimension and $\gamma_{ij}$
is minimal curve in the bulk geometry connecting the boundary
points $x_i$ and $x_j$.

Very interesting issues \cite{Furukawa09}, \cite{Calabrese09}
arise if one regards ${\mathcal A}$ as the union of several
disjoint regions ${\mathcal A} = \cup_i A_i$ and
$\widetilde{\mathcal A}$ as its complement. In the simplest case
one may consider two disjoint blocks $A$ and $B$ such that
${\mathcal A}=A \cup B$. In the analysis of those situations it is
most convenient to compute the mutual information (MI) between
regions $A$ and $B$, which is defined by
\begin{equation}
I_{(A:B)} = S_A+S_B-S_{A\cup B}~.
\label{mutualinf}
\end{equation}

MI measures the amount of correlation (classical and quantum)
between the spatially disconnected regions $A$ and $B$ and acts as
an upper bound on the quantum correlations between operators
defined in those regions \cite{wolf08}; namely,
\begin{equation}
I_{(A:B)} \geq \frac{(\langle {\cal O}_A {\cal O}_B \rangle - \langle {\cal O}_A \rangle \langle {\cal O}_B \rangle  )^2}{2|{\cal O}_A|^2 |{\cal O}_B|^2}~.
\label{wolf_rel}
\end{equation}
The correlators $\langle {\cal O}_A {\cal O}_B \rangle $ as well
as $I_{(A:B)}$ for two spatially disconnected regions disclose
relevant information about the spatial distribution of
entanglement in a given state of the system. However, for two
disjoint blocks, neither the MI nor the quantum correlation
functions happen to be a proper measure of the entanglement since
$A \cup B$ is not a pure state. A true measure of entanglement,
requires the computation of \emph{negativity} \cite{VidalG02}
which is a quite challenging task using field theory methods. \footnote{See, for instance, \cite{Wichterich09_2, Marcovitch09} for a discussion of this issue and some numerical
examples.}

By means of the replica trick, the computation of $I_{(A:B)}$
requires the knowledge of $S_{A \cup B}=-\rm{Tr}(\rho_{A \cup B}
\, \log \rho_{A \cup B})$ with $\rho_{A \cup B} \equiv
\rm{Tr}_{\widetilde{\mathcal A}}\, \rho$, for which very little is
known so far. For two spatially separated regions $A$ and $B$, the
only exact result for $S_{A \cup B}$, has been obtained for free
massless fermions in two dimensions \cite{Casini04, Casini05} but
it remains unknown in its general form for other physically
relevant theories such as the free compactified boson
\cite{Calabrese09}. In a recent paper \cite{Calabrese09}, ${\rm
Tr}\rho_{A\cup B}^n$ for two disjoint intervals $A$ and $B$ - of
length ${\mathit l}$ ($|u_1-v_1|=|u_2 - v_2|={\mathit l}$)
separated by a distance ${\mathit d}=|v_1-u_2|$- has been computed
yielding,
\begin{equation}
{\rm Tr} \rho_{A \cup B}^n = c_n^2 \left(\frac{|u_1-v_1||u_2-v_2|}{|u_1-u_2||v_1-v_2||u_1-v_2||v_1-u_2|} \right)^{\frac{c}6(n-1/n)} {\cal F}_{n}(x)~,
\label{cardycaltonn}
\end{equation}
with $x$ being the conformal four-point ratio defined as
\begin{equation}
x=\frac{|u_1-v_1||u_2-v_2|}{|u_1-u_2||v_1-v_2|}=\frac{{\mathit l}^2}{({\mathit l} + {\mathit d})^2}~.
\label{4pR}
\end{equation}
The function ${\cal F}_{n}(x)$ depends explicitly on the full
operator content of the theory and is, of course, model dependent.
However, the analytic continuation of ${\cal F}_n (x)$ to $n=1$ in
eq. (\ref{cardycaltonn}) is hard to attain and this makes the
computation of $I_{(A:B)}$ between disconnected regions a rather
difficult task within this approach.

In a recent work \cite{headrick10}, using the RT formula for $d=1$
quantum critical system, it has been predicted the occurrence of a
phase transition probed by the computation of the MI between two
disjoint intervals of the boundary CFT$_{d+1}$; namely, as the
conformal four point ratio crosses a critical value the MI
vanishes. Using exact methods, the vanishing of the MI has been
confirmed to occur also for the critical XX spin chain
\cite{korepin11}. This result is quite surprising from a quantum
information point of view since, when the MI vanish, the $\rho_{A
\cup B}$ factorizes into $\rho_{AB}=\rho_{A}\otimes\rho_{B}$,
implying that the two blocks are completely decoupled from each
other and, thus, also the entanglement should be rigorously zero.
In \cite{headrick10} it has been pointed out that,
\begin{equation}
I_{(A:B)}(x) = \begin{cases} 0\,,\quad & x < 1/2 \\ (c/3) \log \left( x /(1-x) \right)\,,\quad & x \geq 1/2\end{cases}\,,
\label{headrickres}
\end{equation}
where $x$ is the conformal four point ratio defined in
(\ref{4pR}). Equation (\ref{headrickres}) states that $I_{(A:B)} =
0$ for $x<1/2$ and it has a discontinuous first derivative at
$x_0=1/2$. As argued in \cite{headrick10}, the discontinuity in
the first derivative of the MI occurs since the shape of the
geodesics (\textit{i.e.}, of the minimal surfaces in the bulk connecting
the two disjoint intervals of the boundary critical system)
changes, as $x$ varies, due to the switching between two saddle
points of the Euclidean action \cite{headrick10, Fursaev06} much similar to the one observed in \cite{hawkingpage82}.

In this paper, inspired by the analysis carried in
\cite{swingle09, swingle10, Evenbly11}, we exploit the holographic structure of the 
\emph{Multiscale Entanglement Renormalization Ansatz} (MERA) tensor networks, to 
analyze the correlations between disjoint blocks of a critical system
described by a $(d+1)$ dimensional conformal field theory lying at
the boundary of an asymptotically AdS$_{d+2}$ spacetime. In order to get an hint on the
pertinent ansatz for the metric to be used, we observe that, when computing the quantum correlations between two disjoint blocks of a boundary quantum critical system,
the structure of the causal cones characteristic of MERA
\cite{Vidal07, VidalG07} implies the existence of two different
regimes attainable by changing a parameter depending on the ratio
between the size and the separation of the disjoint blocks. We
argue that this transition may be accounted by
an AdS$_{d+2}$ black hole geometry and use the RT formula to compute the MI
between the two disjoint regions of the boundary critical system. As an
explicit example, we use a BTZ $\rm{AdS}_3$ black hole to compute
the MI and the quantum correlations between two disjoint intervals
of a one dimensional boundary critical system: here, our analysis
not only confirms the existence of a phase transition emerging
when the conformal four point ratio reaches a critical value but
also provides a rather intuitive entropic argument accounting for
the source of this instability. Finally, we investigate how the
holographic computation of the MI between two disjoint blocks may
be affected by finite size (and temperature) effects. Besides its
appealing beauty, we feel that a remarkable merit of the
holographic approach is that it can help in
establishing fruitful connections between the phase transition
analyzed in \cite{headrick10, Fursaev06} and analogous phase transitions
exhibited by disconnected operators such as the one occurring for
disconnected Wilson loops found in \cite{gross98}.

The paper is organized as follows: in Section 2, we review the
MERA induced AdS/CFT duality \cite{swingle09, Evenbly11} and
analyze its relationship with the RT holographic formula
\cite{Ryu06, Ryu106}; there, we argue that, when considering two
disjoint blocks of the boundary CFT describing the critical
system, the MERA induced AdS/CFT duality leads rather naturally to
the emergence of an AdS black hole as the relevant space time
metric in the dual bulk space. In Section 3 we briefly review the
geometrical properties arising when the space time metric in the
bulk is described by an $\rm{AdS}_3$  BTZ black hole; there we
point out also how a BTZ black hole metric in the MERA induced
dual $\rm{AdS}_3$ space easily accounts for the finite temperature
corrections to the EE. In Section 4 we use the RT formula
\cite{Ryu06, Ryu106} to compute the MI and the quantum
correlations between two disjoint intervals in the $\rm{CFT}_2$
dual to the $\rm{AdS}_3$ BTZ geometry; there we show that the RT
formula, when computed using the $\rm{AdS}_3$ BTZ geometry,
naturally accounts for the phase transition discovered in
\cite{headrick10} and provide an entropic argument accounting for
the emergence of this instability. Finally, in Section 5 we
summarize our results. In the appendix A we use our approach to
compute the MI and quantum correlations between disjoint intervals
of the boundary quantum critical system when the metric of the
MERA induced $\rm{AdS}_3$ space is described by a spinning BTZ
black hole.

\section{MERA induced AdS/CFT duality}\label{MERA_AdSsec}

In \cite{swingle09}, it was firstly observed that MERA
\cite{Vidal07} may give rise to a realization of the AdS/CFT
correspondence \cite{AdSCFTbible00}. This observation has been
subsequently developed in \cite{Evenbly11}. MERA is a real space
renormalization group technique based on a series of consecutive
coarse-graining transformations reducing the amount of
entanglement in a block of lattice sites of a critical system
before truncating its Hilbert space. Namely, by renormalizing the
amount of entanglement in a given system, the MERA procedure
controls the growth of the sites Hilbert space dimension along
successive scaling transformations. This entanglement
renormalization procedure may be encoded in a tensor network
arranged in a set of different levels $\lbrace w_k
\rbrace_{k=0}^{M}$ accounting for the consecutive renormalization
steps and, for quantum systems at criticality, it shows a
characteristic fractal structure. The tensor network implements a
renormalization group transformation which is local in space and
scales local operators into local operators. Furthermore, using
MERA, it has been shown that quantum correlations in the ground
state of one and two dimensional critical quantum many body
systems, could be arranged in layers corresponding to different
length scales \textit{i.e} to different steps in the
renormalization process.

As pointed out in \cite{swingle09, swingle10, Evenbly11}, the
entanglement structure in a quantum critical many body system,
defines a higher dimensional geometry via the renormalization
process described by the scale invariant MERA tensor network. The
emerging geometry can be engineered as follows: all the sites in
the MERA tensor network are arranged in layers, each representing
a different scale (coarse graining renormalization step). As a
result, besides the coordinates labelling the position and the
time $t$, in MERA, one may add a "radial" coordinate $z$ labelling
the hierarchy of scales. Then, the higher dimensional geometry
defined by MERA may be usefully visualized by locating cells
around all the sites of the tensor network representing the
quantum state: these cells are unit cells filling up the emerging
"bulk" geometry and the size of each cell is defined to be
proportional to the entanglement entropy of the site in the cell.
As a result of this procedure a gravity dual picture of the bulk
emerges quite naturally from the entanglement of the degrees of
freedom of the critical system lying on the boundary
\cite{VanRaamsdonk2009}.

The discrete geometry emerging at the critical point is a discrete
version of Anti de Sitter space (AdS) \cite{swingle09, Evenbly11}.
For a one-dimensional quantum critical system with a space
coordinate labelled by $X$, the continuous isometry $w \to w +
\alpha$, $X \to e^{\alpha}X$ of the metric

\begin{equation}\label{ads1}
ds^2\sim dw^2 + e^{-2w}\, dX^2~,
\end{equation}
is replaced by the MERA's discretized version, $w \to w + k$, $X
\to 2^{k}X$ or $X \to 3^{k}X$ depending on the binary or ternary
implementation of the renormalization algorithm \cite{VidalG07}.
The analog of $w=\log{z}$ in the tensor network is simply the
variable labelling the number of renormalization steps carried out
by the MERA algorithm.

When considering a continuous version of MERA \cite{cmera11}, the
discrete AdS-like geometry given by (\ref{ads1}), approaches its
continuous version \textit{i.e} the 3-dimensional AdS space with
the scale invariant metric,

\begin{equation}
ds^2 = \frac{\ell^{2}}{z^{2}}\left(-dt^{2} + dz^2 + dX^2\right)~.
\label{AdS}
\end{equation}
In (\ref{AdS}), $\ell$ is a constant called the AdS radius; it has
the dimension of a length and it is related with the curvature of
the AdS space. With this choice of the space time coordinates the
one dimensional quantum critical system lies at the boundary ($z =
0$) of the bulk geometry.

\subsection{Ryu-Takayanagi formula in the MERA induced AdS/CFT correspondence.}

There is a striking relationship between the RT formula and the
computation of the entanglement entropy in MERA. Using MERA, the
reduced density matrices and hence the quantum correlations are
determined by the structure of the causal cones \cite{Vidal07}.
The causal cone ${\mathcal {CC}}({\mathcal B})$ of a block
${\mathcal B}$ of $l$ sites of the boundary critical system, is
determined by grouping - following all the levels of the MERA
tensor network- all the renormalizing operators and sites which
may affect the sites in the block ${\mathcal B}$. As a result, to
compute the entropy $S_{{\mathcal B}}$ of the block ${\mathcal B}$
it is necessary to trace out any site in the bulk geometry defined
by the tensor network which does not lie in the ${\mathcal
{CC}}({\mathcal B})$ of the block.

The boundary of the ${\mathcal {CC}}({\mathcal B})$ is a curve
$\gamma_{{\mathcal B}}$ in the MERA induced AdS higher dimensional
geometry. The length of $\gamma_{{\mathcal B}}$ is, by definition,
the sum of the entropies of all the traced out sites
\cite{Vidal07, swingle09}, and thus provides an upper bound for
the entropy $S_{{\mathcal B}}$ of ${\mathcal B}$ \cite{Evenbly11},
\begin{equation}\label{entrobound}
S_{{\mathcal B}} \leq \rm{Length}(\gamma_{{\mathcal B}})~.
\end{equation}
The close relationship with the geometrical RT formula comes about
when one realizes that $\gamma_{{\mathcal B}}$ can be regarded as
the minimal curve of RT, since it counts the minimal number of
sites which must be traced out in the coarse graining process.
Indeed, the minimal curve $\gamma_{{\mathcal B}}$ in an optimized
scale invariant MERA network \cite{Evenbly11} has proven to
saturate the bound given in (\ref{entrobound}); this has been
confirmed by explicit computation in one dimensional critical
systems, where it has been shown that $S_{{\mathcal B}} \sim
\frac{c}{3}\log l$ \cite{Vidal07}. Since $\gamma_{{\mathcal B}}$
arises as a boundary of the ${\mathcal {CC}}({\mathcal B})$, it
can be interpreted as an holographic screen which optimally
separates the region in the bulk described by the degrees of
freedom of ${\mathcal B}$, from its complementary region
$\widetilde{\mathcal B}$.

\subsection{Quantum correlations between disjoint blocks from MERA.}

In order to use MERA for computing the two point correlation
functions, one should first observe that, the ${\mathcal {CC}}$s
of two operators located at points $s_1$ and $s_2$ of the boundary
critical system always grow (\textit{i.e} the number of sites
inside a ${\mathcal {CC}}(s_j)$ at MERA level $w_k$, is always
bigger than the number of sites at level $w_{k-1}$), since $k$
increases as one gets deeper into the bulk geometry defined by the
tensor network (\ref{AdS}). As a result, there is a level $w_{*}$
where ${\mathcal {CC}}(s_1)$ and ${\mathcal {CC}}(s_2)$ overlap.
When the ${\mathcal {CC}}$s overlap, the operators defined on the
boundary are correlated with an algebraic decaying functional
dependence \cite{Vidal07}. At variance, the ${\mathcal {CC}}$s of
operators defined on finite size disjoint blocks of the boundary
critical system, tend to exponentially shrink along the
"coordinate" $w$ labelling the MERA level \cite{Vidal07}.

As a result, for two disjoint blocks $A$ and $B$ of the same size
${\it l}$, two situations may occur (see Fig 1) depending only on the distance $d$ between the blocks:\\

$i)$ after $w_H \sim \log {\it l}$ renormalization steps, the ${\mathcal {CC}}(A)$ and the ${\mathcal {CC}}(B)$
shrink to one after they overlap (Fig 1 top). Here one expects that the correlations between the two blocks of the boundary critical system
decay algebraically. \\

$ii)$ after $w_H \sim \log {\it l}$ renormalization steps, the
${\mathcal {CC}}(A)$ and the ${\mathcal {CC}}(B)$ shrink to one
without overlapping (Fig 1 bottom). Here one should expect that
the correlations decay exponentially with the "distance" between the two blocks.\\

It is easy to convince oneself that, if one defines $w_{*} \sim
\log d$,  $i) (ii)$ is realized when $w_H > w_{*}$ ($w_H < w_{*}$).

In the following of this paper, we make the ansatz that an holographic dual spacetime 
that may efficiently account for these two distinct behaviours of the casual cones, 
is given by an AdS$_{d+2}$ black hole geometry of radius $z_H \equiv {\it l}$ ($w_H = \log {\it l}$) when the MI between two disjoint blocks $A$ and $B$ is computed by means of the RT formula.
To support our ansatz we explicitly compute 
the MI and the quantum correlations between disjoint
blocks of a one dimensional quantum critical system using the RT
formula for the AdS$_3$/CFT$_2$ correspondence with the bulk
metric given by a AdS$_3$ BTZ black hole. Under these assumptions,
from equation (\ref{RTarealaw}), the MI reads

\begin{equation}\label{rt_mi}
I_{(A:B)} = \frac{1}{4 G_{N}^{(3)}} \left[ {\rm Length}(\gamma_A) + {\rm Length}(\gamma_B) - {\rm Length}(\gamma_{A \cup B}) \right]~,
\end{equation}
with $\gamma_A$, $\gamma_B$ and $\gamma_{A \cup B}$ being geodesic
curves in the BTZ black hole background \cite{btz92}. As we shall
see in the following sections, a computation of the MI carried
using this approach supports- from a different point of view- the
results obtained in \cite{headrick10}.
\begin{figure}
\includegraphics[width=4.0 in]{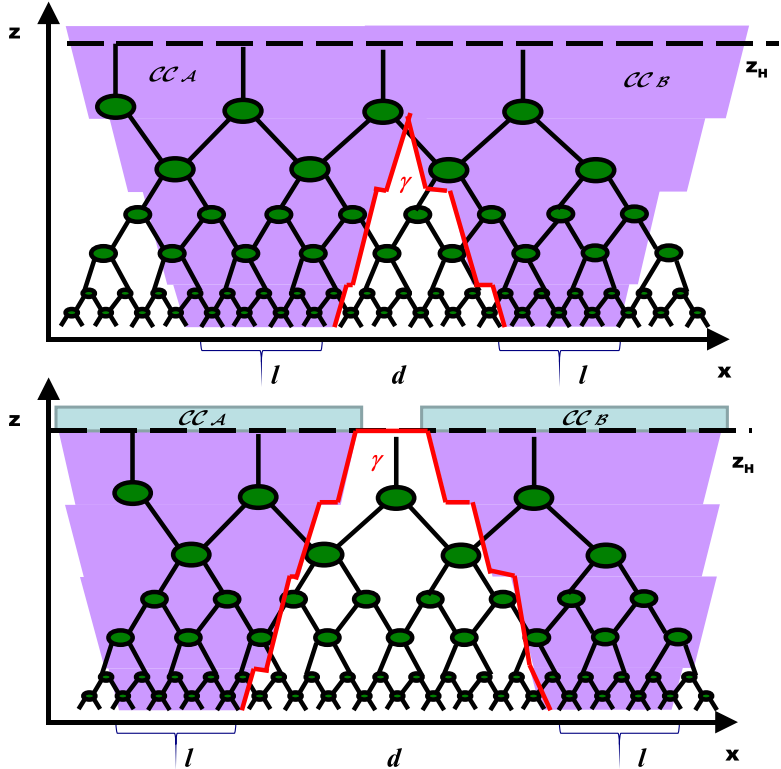}
\label{Fig1} \centering \caption{{\bf Top}: Schematic
representation of MERA ${\mathcal {CC}}$ for two disjoint finite
intervals $A$ and $B$ when the separation between them allows for
overlapping after $w_{*} \sim \log {\mathit d}$ renormalization
steps. The overlap occurs before the causal cones shrink to one
(in our representation, when causal cones stabilize their width
after $w_H = \log {\it l}$ renormalization steps). {\bf Bottom}:
Schematic representation of MERA ${\mathcal {CC}}$ for two
disjoint finite intervals $A$ and $B$ when the separation between
them does not allow for overlapping after $w^* \sim \log {\mathit
d}$ renormalization steps. The curve $\gamma$ that goes through 
the links between the nodes of the MERA network is the minimal curve
separating the ${\mathcal {CC}}(A)$ and ${\mathcal {CC}}(B)$ from
the traced out sites in the MERA bulk geometry (\ref{AdS}).}
\end{figure}

\section{The BTZ black hole}\label{BTZsec}
\subsection{BTZ black hole solution}

Ba{\~n}ados, Teitelboim, and Zanelli (BTZ) showed that
(2+1)-dimensional gravity has a black hole solution, the BTZ black
hole, differing from the Schwarzschild and Kerr solutions mainly
in that it is asymptotically anti-de Sitter rather than
asymptotically flat. The BTZ solution is clearly a black hole: it
has an event horizon and (when rotating) an inner horizon, and it
exhibits thermodynamic properties much like those of a
(3+1)-dimensional black hole \cite{btz92}.

The BTZ black hole may be obtained by orbifolding $\rm{AdS}_3$
through $SL(2,{\mathbf{C}})$ identifications \cite{kraus06} and is
a solution of pure gravity in three dimensions with a negative
cosmological constant described by the Einstein-Hilbert action
supplemented by boundary terms \cite{kraus06},

\begin{equation}
I =  \frac{1}{16\pi G } \int \! d^3 x \sqrt{g} \,(R-\frac{2}{\ell^2}) +I_{\rm bndy}~.
\label{action}
\end{equation}

In the following we use the Euclidean signature, and use the
notation of Misner, Thorne, and Wheeler \cite{gravitation}; as a
result, $r \equiv 1/z$ so that the boundary is now located at $r
\to \infty$. A simple solution of the equations of motion is just
the $\rm{AdS}_{3}$ spacetime,

\begin{equation}
ds^2 = (1+r^2/\ell^2)dt^2 +\frac{dr^2}{ 1+r^2/\ell^2}+r^2 d\phi^2~.
\label{AdSr}
\end{equation}
$\rm{AdS}_{3}$ has maximal symmetry, with the isometry group being
$SL(2,{\mathbf{C}}) \cong SL(2,{\mathbf{R}})_{L} \times
SL(2,{\mathbf{R}})_{R}$.

A more general one-parameter family of solutions is provided by
the non-rotating BTZ black hole of mass $M$ \cite{kraus06},

\begin{equation}
ds^2 = \frac{(r^2- r_+^2)}{\ell^2}dt^2 + \frac{\ell^2}{(r^2-r_+^2)} dr^2 + r^2 d\phi^2~,
\label{nonrotBTZ}
\end{equation}
describing an $\rm{AdS}$ black hole with an event horizon located
at $r=r_{+}=\ell \sqrt{8 G M}$ at a temperature $T= r_{+}/2 \pi
\ell^2$; of course, for large $r$, the solution (\ref{nonrotBTZ})
asymptotically approaches $\rm{AdS}_{3}$.

The metric of a rotating BTZ black hole of mass $M$ and angular
momentum $J$ is given, instead, by
\begin{equation}
 ds^2 = \frac{(r^2 -r_+^2)(r^2 -r_-^2)}{r^2
\ell^2}dt^2+\frac{\ell^2 r^2}{
(r^2-r_+^2)(r^2-r_-^2)}dr^2+r^2(d\phi + i\frac{r_+  r_-}{\ell r^2} dt)^2~.
\label{rotBTZ}
\end{equation}
with
\begin{eqnarray}
M = \frac{r_{+}^2 + r_{-}^2}{\ell^2}~, \quad J =  \frac{2 r_{+} r_{-}}{\ell}~, \\
r_{\pm} = \ell \left[\frac{M}{2}\left(1 \pm \sqrt{1-\left( \frac{J}{M \ell}\right)^2 }\right)  \right]^{\frac{1}{2}} ~.
\label{paramsbtz}
\end{eqnarray}
Now, the event horizon is located at $r=r_{+}$  with $r_{+} \geq
r_{-}$ and $r_{-}$ being the inner Cauchy horizon. Rotating BTZ
black holes have been recently shown to be relevant in
investigations of helical Tomonaga-Luttinger liquids
\cite{Balasu2010}.

\subsection{Dual CFT to the BTZ solution}\label{cftsec}

The boundary of asymptotically $\rm{AdS}_{3}$ spacetimes is a two
dimensional torus on which one can define a dual CFT with its
conformal symmetry being generated by two copies of the Virasoro
algebra acting separately on the left and right moving sectors. As
a result, the CFT splits into two independent sectors at thermal
equilibrium with temperatures,
\begin{eqnarray}
T _{L}= \frac{r_{+} + r_{-}}{2 \pi \ell^2}~, \quad T_{R} =  \frac{r_{+}- r_{-}}{2 \pi \ell^2}~.
\label{temperatures}
\end{eqnarray}
The mass $M$ and the angular momentum $J$ in the rotating BTZ
black hole geometry are related to the  Virasoro charges of the
dual CFT on the boundary by
\begin{eqnarray}
L_0 -\frac{c}{24} = \frac{1}{16 G} (M\ell + J)~,\quad
\widetilde{L}_0-\frac{\widetilde{c}}{24} = \frac{1}{16 G}(M\ell - J)~.
\label{virgen}
\end{eqnarray}
with $c = \widetilde{c}$ given by the Brown-Henneaux holographic
relation \cite{brownhenn}

\begin{equation}
c = \frac{3 \ell}{2 G_N^{(3)}}~.
\label{cch}
\end{equation}

In the non rotating BTZ metric (\ref{nonrotBTZ}) one has
\begin{eqnarray}
L_0 -\frac{c}{24} = \widetilde{L}_0
-\frac{\widetilde{c}}{24} = \frac{r_{+}^2}{16 G \ell}~.
\label{virgen2}
\end{eqnarray}
It is easy to prove \cite{kraus06} that, for the AdS$_3$ metric,
(\ref{AdSr}) $L_0  = \widetilde{L}_0=0$; this is just a
consequence of its invariance under the $SL(2,\mathbf{R})_L \times
SL(2,\mathbf{R})_R $ group of isometries generated by $L_{0,\pm
1}$ and  $\widetilde{L}_{0,\pm 1}$.

\subsection{Geodesics in the BTZ geometry}

The RT formula uses the spacelike geodesics in a given metric. For
the BTZ black hole these geodesics are well known. The length of
the geodesics connecting two points $x_{i}$ and $x_{j}$ separated
by a distance $\vert x_{i} - x_{j}\vert$ and located at the
boundary of the AdS$_3$ space whose metric is described by a BTZ
Black Hole (\ref{nonrotBTZ}), can be written as \cite{Ross00},
\begin{equation}
\mathit{L}(x_{i},x_{j}) =  2 \ell \log\left[\frac{\beta}{\pi \epsilon} \sinh\left(\frac{\pi \vert x_{i} - x_{j}\vert}{\beta} \right)  \right]~,
\label{spinlessbtzgeo}
\end{equation}
with $\beta = 2 \pi \ell^2/r_{+}$ and $\epsilon$ the regularizing
boundary cut-off.

Using the RT formula (\ref{RTarealaw}) and (\ref{cch}), one gets
the well known formula \cite{Calabrese04} for the EE of a single
connected block $A$ of length $\ell_{A}=\vert x_{1} - x_{2}\vert$
from the BTZ geometry; namely, one has that

\begin{eqnarray}\label{EE1nterval}
S_{A}=\frac{\mathit{L}(x_{1},x_{2})}{4G_N^{(3)}}=\frac{c}{3} \log\left[\frac{\beta}{\pi \epsilon} \sinh\left(\frac{\pi \ell_{A}}{\beta} \right)  \right] \nonumber \\
\approx  \begin{cases}
(c/3) \log \left(\ell_{A}/\epsilon \right) \,,\quad & r_{+} \to 0\   (\beta \to \infty)\\
\\
(\pi c/3)(\ell_{A}/\beta) + \frac{c}{3}\log (\beta/2 \pi \epsilon)\,,\quad & r_{+} \to \infty\ (\beta \to 0)
\end{cases}\,.
\end{eqnarray}
As expected, from eq. (\ref{EE1nterval}) one recovers the
logarithmic dependence only in the zero temperature limit (\textit{i.e.},
when the size of the interval is small in comparison with the
distance of the horizon from the boundary); indeed, in this limit,
the BTZ geodesics stay close to the boundary and only probe the
asymptotic form of the ${\rm AdS}_3$ BTZ geometry. (Fig.2 Left).
At variance, when the size of the simply connected region $A$ is
bigger than the distance of the horizon from the boundary, the BTZ
geodesics probe the black hole horizon extending tangentially to
it; this induces the linear correction to the EE which, for a
single connected region $A$, describes now a thermal state at
temperature $T=1/\beta$ (Fig.2 Right).

For a rotating black hole, the geodesics are given instead by
\begin{equation}
\mathit{L}(x_{i},x_{j}) =  2 \ell \log\left[\frac{\beta_{L} \beta_{R}}{\pi^2 \epsilon^2} \sinh\left(\frac{\pi \vert x_{i} - x_{j}\vert}{\beta_{L}} \right) \sinh\left(\frac{\pi \vert x_{i} - x_{j}\vert}{\beta_{R}} \right) \right]~,
\label{spinbtzgeo}
\end{equation}
where $\beta_{L,R} = 1/T_{L,R}$. As a result one gets that
\begin{align}
S_{A}= S_{A}^{L} + S_{A}^{R}=  \nonumber \\
\nonumber \\
=\frac{c}{3} \log\left[\frac{\beta_{L}}{\pi \epsilon} \sinh\left(\frac{\pi \ell_{A}}{\beta_{L}} \right)\right] + \frac{c}{3} \log\left[\frac{\beta_{R}}{\pi \epsilon} \sinh\left(\frac{\pi \ell_{A}}{\beta_{R}} \right) \right]~.
\label{EE1intervalspin}
\end{align}
Equation (\ref{EE1intervalspin}) factorizes into left and right
moving sectors as expected from the left-right decoupling of the
CFT$_2$.

\begin{figure}
\includegraphics[width=5.5 in]{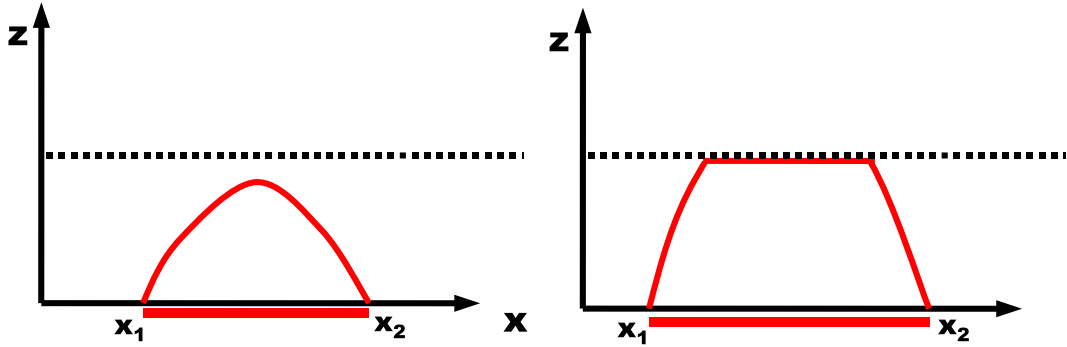}
\label{Fig2} \centering \caption{Geodesic used in computing the
entanglement entropy of a region of length ${\mathit l}=\vert
x_1-x_2\vert$ in the AdS Black Hole geometry. {\bf Left}: $\gamma$
does not approach the horizon (dotted line). {\bf Right}: $\gamma$
wraps around the horizon.}
\end{figure}

\section{Holographic Computation of Quantum Correlations and Mutual Information for two disjoint intervals}\label{resec}

In this section we derive both the MI and the quantum correlations
between two disjoint intervals of a one dimensional critical
system described by a CFT$_2$ located at the boundary of the
AdS$_3$ space. We assume in the following that the disjoint
intervals $A$ and $B$ have equal size ${\mathit l}$ and are
separated by a distance ${\mathit d}$. Namely, we take $A \equiv
\left[u_1, v_1 \right], B \equiv \left[u_2,v_2 \right]$ with $|u_1
- v_1|=|u_2 - v_2| = {\mathit l}$ and $|v_1- u_2|={\mathit d}$
(see Fig.3). We shall see how, in both computations, one can find
a critical value of a pertinent parameter at which there is a
transition between two very different behaviors.

For the two disjoint intervals $A$ and $B$ the holographic
computation of the MI requires to determine the minimal curve in
the bulk homologous to $A \cup B$. In the MERA induced AdS/CFT
correspondence, the curve $\gamma_{A \cup B}$ is generated by
tracing out the bulk sites lying outside the ${\mathcal {CC}}(A)$
and ${\mathcal {CC}}(B)$ and is an holographic screen for the entropy contained in
 $A \cup B$. As a result, for generating
this holographic screen, there are- just as in \cite{headrick10}-
only two possible options given by $\gamma_{A \cup B}^{(con)}$
(Fig 3 Left) and $\gamma_{A \cup B}^{(dis)}$ (Fig 3 Right),
respectively: the curve $\gamma_{A \cup B}^{(con)}$ ($\gamma_{A
\cup B}^{(dis)}$) corresponds to the overlapping (non-overlapping)
configuration of the causal cones ${\mathcal {CC}}(A)$ and
${\mathcal {CC}}(B)$ depicted in Fig 1. Namely, $\gamma^{(dis)}_{A
\cup B}$, describes a situation in which the two intervals are
enough separated so that ${\mathit L}_{A \cup B}={\mathit
L}_1(u_1,v_1) + {\mathit L}_2(u_2, v_2)$, while $\gamma^{(con)}_{A
\cup B}$, describes a situation where the separation between the
intervals is so small that the minimal curve of the region $A \cup
B$, connects the inner and outer boundaries of the two regions so
that ${\mathit L}_{A \cup B}={\mathit L}_1(u_1,v_2) + {\mathit
L}_2(v_1, u_2)$.

Of course, when $\gamma^{(dis)}_{A \cup B}$ is used in the
holographic computation of the MI, the MI vanishes as a
consequence of eqs. (\ref{RTarealaw}) and (\ref{rt_mi}). At
variance, when one uses $\gamma_{A \cup B}^{(con)}$, the
holographic computation of the MI (\ref{rt_mi}), using as the
metric of the AdS${_3}$ bulk space the one corresponding to a BTZ
black hole with the horizon located at $z_+={\mathit l}$ from the
boundary \textit{i.e} $\beta=2 \pi z_+$, yields (\ref{cch})
\begin{equation}
\label{minonrot}
I_{(A:B)}=\frac{c}{3}\log\left[\frac{\sinh\left(\pi T \vert u_1 - v_1 \vert
\right)\, \sinh\left(\pi T \vert u_2 - v_2 \vert
\right)}{\sinh\left(\pi T \vert u_1 - v_2 \vert \right)\, \sinh\left(\pi T
\vert v_1 - u_2 \vert  \right)} \right]~,
\end{equation}
with $T=1/\beta$.

One sees that $I_{(A:B)}$ in eq. (\ref{minonrot}) equals zero when
a certain ratio between ${\mathit l}=|u_1 - v_1|=|u_2 - v_2|$ and
${\mathit d=|v_1- u_2|}$ is reached; namely, one sees that the MI, when computed
using the BTZ black hole as the metric of AdS$_3$, vanishes at a value
of the conformal four point ratio given by $x_0 \sim 0.53$.
This is in agreement with the result of
\cite{headrick10}. However, an advantage of the MERA induced
AdS/CFT correspondence lies in the fact that one can provide a
rather intuitive entropic argument accounting for the use of either
one of the two minimal curves depicted in Fig.3 when performing
the holographic computation of MI. Indeed, since the length of the
curves $\gamma_{A \cup B}^{(con)}$ and $\gamma_{A \cup B}^{(dis)}$
are- by definition- the sum of the entropies of all the traced out
sites, the transition between the two behaviors of MI
occurs when the separation between the two (equal size)
disjoint blocks $A$ and $B$ is such that the entropy due to the
the tracing out process yielding $\gamma_{A \cup B}^{(con)}$
equals the
entropy due to the tracing out process yielding $\gamma_{A \cup B}^{(dis)}$. \\

A similar transition is found also in the computation of the
quantum correlations between two primary operators ${\cal O}(x_A)$
and ${\cal O}(x_B)$ ($x_A \in A$ and $x_B \in B$) with conformal
dimension $\Delta$ defined in the CFT$_d+1$ describing the
boundary critical system. This should be expected in view of the
bound (\ref{wolf_rel}). The $\rm{AdS}/\rm{CFT}$ correspondence
implies \cite{AdSCFTbible00, Witten98} that
\begin{equation}
\langle {\cal O}(x_A) {\cal O}(x_B) \rangle \sim e^{-m {\mathit
L}(x_A,x_B)}~, \label{2pop}
\end{equation}
where $\Delta \approx m \ell$ and ${\mathit L}(x_A,x_B)$ is the
length of the shortest geodesic connecting the boundary points
$x_A$ and $x_B$. Using for ${\mathit L}(x_A,x_B)$ the expression
given in (\ref{spinlessbtzgeo}), one easily gets,
\begin{eqnarray}
\langle {\cal O}(x_A) {\cal O}(x_B) \rangle \sim \left[\frac{\pi T}{\sinh \left(\pi T |x_A - x_B|\right)}   \right]^{2\Delta}\nonumber \\
\approx \begin{cases}
|x_A-x_B|^{-2\Delta} \,,\quad & z_{+} \gg |x_A-x_B|\   \\
\\
z_+^{-2\Delta} \exp\left( -2 \pi T \Delta |x_A - x_B| \right)
\,,\quad & z_{+} \ll |x_A-x_B|\
\end{cases}\,.
\label{adscorr}
\end{eqnarray}
From (\ref{adscorr}) one sees that there is a change from an
algebraic to an exponential decaying behavior of the two point
quantum correlation function and that the transition between the
two regimes occurs when $\langle {\cal O}(x_A) {\cal O}(x_B)
\rangle \sim e^{-\Delta}$. When this happens, one has that
\begin{equation}
2 \pi\, T\,  \vert x_A - x_B \vert = \frac{\vert x_A - x_B \vert}
{z_{+}} \approx 1~,
\end{equation}
which defines the value of the parameter $\mu = \vert x_A - x_B
\vert/{\mathit l}$ at which this transition occurs \cite{tonni10}.

\begin{figure}
\includegraphics[width=5.5 in]{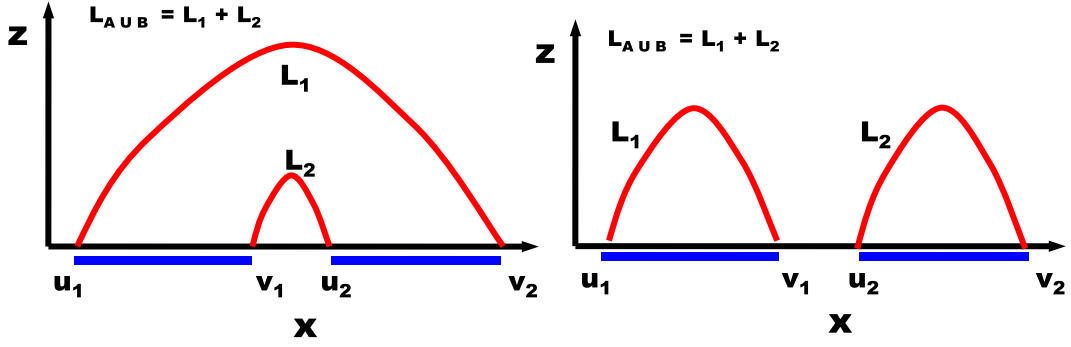}
\label{Fig3} \centering \caption{Minimal curves used in the
holographic computation of $S_{A \cup B}$ and $I_{(A:B)}$ for to
disjoint intervals $A$ and $B$.}
\end{figure}

Eqs. (\ref{minonrot}) and (\ref{adscorr}) are derived for infinite
systems when the central charge $c \to \infty$. However, one may
be interested in the behavior of the MI and of the quantum
correlations in a regime where both the temperature $T$ and the
size of the system $L$ are finite \cite{Birmingham2002}. In
particular, one is interested in knowing if the transition between
the two very distinct behaviors found for the infinite system is
still attainable and, if so, how the critical value of the
pertinent parameter is going to be affected when $T$ and $L$ are
finite. To grasp how the results obtained so far in this section
are going to be changed due to these finite size effects we look
at the behavior of the two point correlation functions of free
fermions on the torus \cite{Francesco97}. For this system one has
that,

\begin{equation}
\langle\psi (u)\psi(v)\rangle_{\nu}=\frac{\theta_{\nu} (i|u - v|T\, |\tau)}{\theta_{\nu}(0|\tau)}\, \frac{\partial_{\omega}\theta_{1}(0|\tau)}{\theta_1 (i|u - v|T\, |\tau)}~,
\label{ffcorr}
\end{equation}

where $\theta_{\nu}(\omega \vert \tau)$ are the modular Jacobi theta functions \cite{tatatheta}, 
$\partial_{\omega}\theta_{1}(0|\tau) \equiv \partial_{\omega}\theta_{1}(\omega|\tau)\vert_{\omega=0}$,  
$\nu$ defines the boundary conditions for $\psi$ and $\tau \equiv i L T$. For
instance, for finite temperature boundary conditions, only the
sectors $\nu=3,4$ of the spin structure of the fermion contribute
(\ref{ffcorr}); this is to say that, on the torus, one can only
choose for $\psi$ either $\nu=3$, corresponding to
antiperiodic-periodic (Neveu-Schwarz, NS - Ramond, R) boundary
conditions, or $\nu=4$ which corresponds to
antiperiodic-antiperiodic (NS-NS) boundary conditions.

When $L T \to \infty$, using the standard representation of the
$\theta_{\nu}$ functions \cite{tatatheta, Alvarez86}, one gets

\begin{equation}
\langle\psi (u)\psi(v)\rangle_{3(4)}=\frac{\pi T}{4\sinh{\pi T |u - v|}}[1\pm 2e^{-\pi L T}\cosh 2\pi T |u - v|+...]~,
\label{theta_app}
\end{equation}
As a result, in the limit of finite $T$ with $L
\gg |u - v|$, one may approximate Eq. (\ref{adscorr}) with $\Delta =
1/2$ in terms of (\ref{ffcorr}) and write (\ref{minonrot}) as,

\begin{equation}
I_{(A:B)}=\frac{c}{3} \left[\frac{\Upsilon(v_2,u_1)\, \Upsilon(u_2,v_1)}{\Upsilon(v_1,u_1)\, \Upsilon(v_2,u_2)} \right] + \frac{c}{3}\log\left[\frac{\theta_{\nu} (i|u_1 - v_2|T\, |\tau)\, \theta_{\nu} (i|u_2 - v_1|T\, |\tau)}{\theta_{\nu} (i|u_1 - v_1|T\, |\tau)\, \theta_{\nu} (i|u_2 - v_2|T\, |\tau)} \right]~,
\end{equation}
where $\Upsilon(u,v)$ is given
by \cite{Francesco97}

\begin{equation}
\Upsilon(u,v)= \log \frac{\partial_{\omega}\theta_{1}(0|\tau)}{\theta_1 (i|u - v|T\, |\tau)}~.
\end{equation}
For $L\gg |u - v|$ one has that $\Upsilon(u,v) \sim \log 1/(i|u - v| T)$; as a
result one has

\begin{equation}
I_{(A:B)}=\frac{c}{3}\log\left(\frac{x}{1-x} \right)+ \frac{c}{3}\log\left[\frac{\theta_{\nu} (i|u_1 - v_2|T\, |\tau)\, \theta_{\nu} (i|u_2 - v_1|T\, |\tau)}{\theta_{\nu} (i|u_1 - v_1|T\, |\tau)\, \theta_{\nu} (i|u_2 - v_2|T\, |\tau)} \right] ~,
\label{mitheta}
\end{equation}
where $x$ is defined in (\ref{4pR}).
One notices that (\ref{mitheta}) reduces to
(\ref{headrickres}) when the separation between the intervals
is very small since the function $f_{\nu}(x,\tau)$ defined as

\begin{equation}
f_{\nu}(x,\tau)=\log \left( \frac{\theta_{\nu} (i|u_1 - v_2|T\, |\tau)\, \theta_{\nu} (i|u_2 - v_1|T\, |\tau)}{\theta_{\nu} (i|u_1 - v_1|T\, |\tau)\, \theta_{\nu} (i|u_2 - v_2|T\, |\tau)}\right)~,
\label{bhcorrt}
\end{equation}
approaches zero when $x \to 1$ while $f_{\nu}(x,\tau)>0$ for $x \leq 1$.

We observe that, in a rather large range of values for $T$
and $L$, there is still a transition between two very different
behaviors of the MI. However, the critical value $x_0$, at which
the transition occurs strongly depends on the ratio ${\mathit
l}/L$ \textit{i.e}, $1/|\tau| = (LT)^{-1}$ as reported in Fig. 4.
Indeed, a numerical analysis shows that, for a finite system,
$x_0$ is always $x_0 < 1/2$ and that, only as $L \to \infty$, $x_0
\to 1/2$ recovering the result in \cite{headrick10}.

\begin{figure}
\includegraphics[totalheight = 3.5 in]{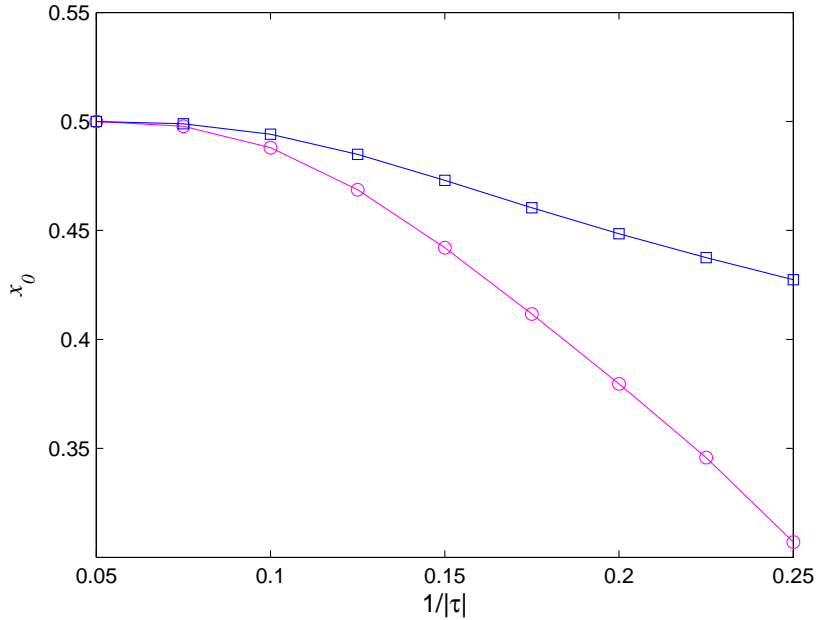}
\label{Fig4} \centering \caption{ Dependence of the transition
point $x_0$ on $|\tau|= LT $ \textit{i.e} $1/|\tau| \propto
{\mathit l}/L$ for the non rotating BTZ black hole (Eq
(\ref{mitheta}), circles) and for the quasi-extremal rotating BTZ
black hole (Eq (\ref{mitheta2}), squares).}
\end{figure}

For the sake of completeness we shall compute the quantum
correlators and the MI in a rotating BTZ black hole background in
Appendix A.

 \section{Concluding Remarks}

Originally developed within string theory, the AdS/CFT
correspondence provides a geometrical framework to investigate
also strongly coupled condensed matter and spin systems at
criticality . An intriguing observation has been that MERA
\cite{Vidal07} may be efficiently described through the AdS/CFT
correspondence by introducing an AdS metric in a pertinently
engineered bulk space \cite{swingle09, Evenbly11}. In this paper we
use this MERA induced AdS/CFT correspondence to provide a
framework in which the mutual information and the two point
quantum correlations between disjoint blocks of a quantum system
at criticality may be evaluated. We feel that an advantage of this
approach is that, at least in principle, is not strictly confined
to the analysis of one dimensional critical systems.

In order to get an hint on the pertinent metric to be used to
describe the MERA induced bulk AdS$_{d+2}$ space, we observed here
that, when computing the quantum correlations between two disjoint
blocks of a boundary quantum critical system, the structure of the
causal cones characteristic of MERA implies the existence of two
different regimes attainable by tuning the ratio between the size
and the separation of the disjoint blocks. To account for this
transition we proposed that the MERA induced holographic dual bulk
spacetime could be described by an AdS$_{d+2}$ black hole and
used the RT formula to compute the MI of two disjoint regions of
the boundary critical system. Intuitively speaking, this amounts
to orbifolding the AdS geometry introduced in \cite{swingle09,
Evenbly11} when dealing with disjoint blocks.

As an explicit example, we used a BTZ $\rm{AdS}_3$ black hole to
compute the MI and the quantum correlations between two disjoint
intervals of a one dimensional boundary quantum critical system:
here, our analysis not only confirmed the existence of the phase
transition emerging when the conformal four point ratio reaches a
critical value but also provided a rather intuitive entropic
argument explaining the source of this instability. Furthermore,
we showed how non universal behaviors may emerge in
the holographic computation of the MI between two well separated
disjoint blocks. Of course, our analysis does not exclude the possibility 
that other geometries -such as Lifshitz geometries- may account for the 
behavior of the causal cones of disjoint blocks in MERA.

A remarkable feature of the RT approach to the computation of MI and EE 
taken in this paper is that it associates with each spatial region of the boundary
a unique spatial region of the bulk \cite{Fursaev06}. This bulk to boundary map -via the structure of the causal cones- seems to play an intriguing role also in MERA.
Indeed, we exploited this map in MERA to give an ansatz for the dual holographic geometry associated to a region made of two disjoint blocks of the $d$-dimensional boundary critical system. We, then, observed that -when the separation between the two blocks exceeds a critical
value- the quantum correlations exhibit a thermal behaviour and the EE may be computed as the thermodynamic entropy associated to a certain black hole. In the context of the AdS/CFT correspondence thermal states have been recently constructed in \cite{CasHuMyers2011}.
 
The AdS/CFT correspondence, is a strong-weak duality. This amounts
to say that, when the dual gravity description of a quantum system
is classical, the correlations on the boundary theory are quantum
and highly non-local (entanglement) \cite{VanRaamsdonk2009}. In
the MERA induced AdS/CFT correspondence, the locality of the
emerging AdS space is due to the existence of entanglement at all
scales in the quantum critical system located at the boundary. We
feel that our results may help to elucidate the nature (quantum
and/or classical) of the correlations computed using the RT
formula within the AdS/CFT correspondence. Indeed, despite the
fact that MI quantifies both classical and quantum correlations,
recently, in \cite{headrick11}, it has been shown that MI, when
computed using the holographic RT formula, obeys the same monogamy
relations required for a true measure of entanglement. Since the
monogamy relations severely limit the amount of entanglement
sharable between the different parts of an arbitrarily partitioned
system \cite{coffman00} this should imply a truly quantum nature
of the correlations measured in the holographic computation of the
MI. In \cite{Wichterich09_2}, using numerical methods to compute a
true measure of entanglement such as negativity, it has been found
that the entanglement between disjoint intervals in spin chains at
criticality also showed a crossover from pure algebraic decay to
pure exponential decay when a critical ratio between the
separation and the size of the intervals was reached.

Finally, we feel that the use of a pertinent metric in the AdS space built from the MERA
induced AdS/CFT correspondence may be exploited also as a way to
look for alternative and- hopefully- more powerful ways of
optimizing MERA tensor networks.

\appendix
\section{Quantum correlators and MI in the rotating BTZ background}

In this  appendix we compute the MI and quantum correlations
between two disjoint intervals of the same size ${\mathit l}$ when the
background metric is a
rotating BTZ black hole.

When the distance between the two intervals is small enough, the
geodesic of minimal length is ${\mathit L}_{A \cup
B}^{(con)}={\mathit L}(u_1,v_2) + {\mathit L}(v_1, u_2)$; using
(\ref{spinbtzgeo}), one gets
\begin{equation}
I_{(A:B)}=\frac{c}{3} \log\left[\frac{\sinh^2\left(\pi T_L \varrho_1 \right) }{\sinh\left(\pi T_L \varrho_2\right) \sinh\left(\pi T_L \varrho_3 \right)} \frac{\sinh^2\left(\pi T_R \varrho_1 \right) }{\sinh\left(\pi T_R \varrho_2 \right) \sinh\left(\pi T_R \varrho_3\right)} \right]~,
\label{mutinforot}
\end{equation}
where $\varrho_1\equiv |u_1 - v_1|=|u_2 - v_2|={\mathit l}$, $\varrho_2 \equiv |u_2 - v_1|={\mathit d}$ and $\varrho_3 \equiv |u_1 - v_2|= 2{\mathit l} + {\mathit d}$. In equation (\ref{mutinforot}), the event horizon is located at
$z_+ = 1/r_+={\mathit l}$ and, upon introducing the two variables
$z_L=1/(r_{+} +r_{-})$ and $z_R=1/(r_{+} -r_{-})$ such that
$z_L<z_+<z_R$, one is able to define the two temperatures $T_L=1/2
\pi z_L$ and $T_R=1/2 \pi z_R$ (Fig.5).

For the near extremal BTZ black hole, \textit{i.e}, for a black
hole in which $M \ell \gtrsim J$ ($r_+ \gtrsim r_{-}$, $z_R \to
\infty$), equation (\ref{mutinforot}) may be written as,
\begin{equation}
I_{(A:B)}=\frac{c}{3} \log\left(\frac{x}{1-x}\right) + \frac{c}{3} \log\left[\frac{\sinh^2\left(\pi T_L \varrho_1 \right)} {\sinh\left(\pi T_L (\pi T_L \varrho_2 \right) \sinh\left(\pi T_L \varrho_3 \right)} \right]~.
\label{mibtz}
\end{equation}
One sees from equation (\ref{mibtz}) that the decoupling of the
right and left moving sectors induced by the presence of two
horizons, plus the near extremality condition of the spinning
black hole, yields an expression for the MI which decomposes in
two terms: one -identical to (\ref{headrickres})-depends only on
$c$ and the conformal ratio $x$- and a second identical to
(\ref{minonrot}). It is easy to convince oneself that equation
(\ref{mitheta}) when $T=T_L$ reproduces the second term of
(\ref{mibtz}). As a result, the MI of two disjoint intervals in a
finite system of total length $L$ may be written as,

\begin{equation}
I_{(A:B)}=\frac{2 c}{3} \log\left(\frac{x}{1-x}\right) + \frac{c}{3}\log \left( \frac{\theta_{\nu} (i|u_1 - v_2|T_L\, |\tau)\, \theta_{\nu} (i|u_2 - v_1|T_L\, |\tau)}{\theta_{\nu} (i|u_1 - v_1|T_L\, |\tau)\, \theta_{\nu} (i|u_2 - v_2|T_L\, |\tau)}\right) ~,
\label{mitheta2}
\end{equation}
with $\tau = i L T_L$.

For the rotating extremal BTZ black hole, numerical simulations
show that $1/|\tau|$ weakly affects the location of the transition
point $x_0$ for the MI (Fig. 4). Furthermore, it appears from our
results and those presented in \cite{headrick10}, that the MI is
parametrically small at $x_0$. As a result, due to the inequality
(\ref{wolf_rel}), one should expect also here a transition for the
quantum correlations.

\begin{figure}
\includegraphics[width=4 in]{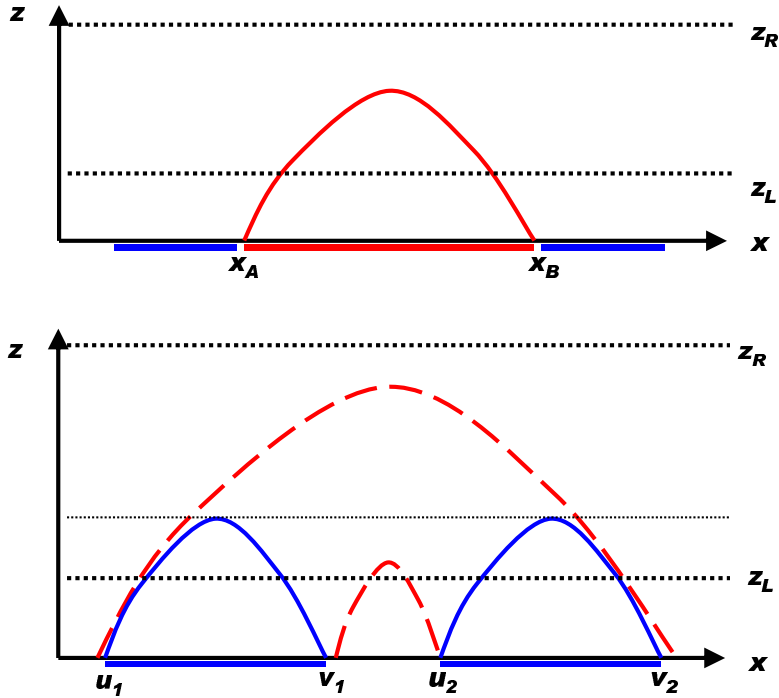}
\label{Fig5} \centering \caption{Characteristic length scales
$z_R$ and $z_L$ of the rotating BTZ black hole background. The
event horizon $z_+$ lies between $z_L<z_+<z_R$ (thin dotted line
between $z_L$ and $z_R$). {\bf Top}. Geodesic connecting the two
closest boundary points of the intervals $A$ and $B$ used in the
computation of eq. (\ref{2pcspin}). {\bf Bottom}: Geodesic
${\mathit L}_{A \cup B}^{(con)}={\mathit L}(u_2, v_2) + {\mathit
L}(v_1,u_2)$ used in the computation of eq. (\ref{mibtz}).}
\end{figure}

For the rotating BTZ black hole, the behavior of the quantum
correlations between points located in different disjoint
intervals is given by,
\begin{equation}
\langle {\cal O}(x_A) {\cal O}(x_B) \rangle \sim \left[\frac{\beta_L \beta_R}{\pi^2} \sinh \left(\frac{\pi |x_A - x_B|}{\beta_L}\right) \sinh \left(\frac{\pi |x_A - x_B|}{\beta_R}\right)  \right]^{-2\Delta}~.
\label{2pcspin}
\end{equation}

When the near extremality condition is satisfied, ${\it i.e}$,
when $r_+ \gtrsim r_{-}$, so $\beta_R \to \infty$, the two point
correlations behave as,
\begin{equation}
\langle {\cal O}(x_A) {\cal O}(x_B) \rangle \sim \left( \frac{1}{|x_A-x_B|}\right)^{2\Delta} \left( \frac{2 \pi}{\beta_L}\right)^{2\Delta} \exp\left(- \frac{2 \pi \Delta |x_A-x_B|}{\beta_L} \right)~.
\label{nearextr2p}
\end{equation}

Eq. (\ref{nearextr2p}) shows the existence of a crossover from
pure algebraic decay to pure exponential decay of the quantum
correlations  as $\mu = \vert x_A - x_B\vert / z_{+}$ increases.

\begin{acknowledgments}

We are grateful to S. Bose and H. Wichterich for many very
fruitful insights and stimulating correspondence at the early
stages of this project. We thank J. Hung and G. Grignani for a
critical reading of the manuscript. We benefited from discussions
with R.C. Myers, J.I Cirac, R.N.C Pfeifer, G. Evenbly, J. McGreevy
and B. Swingle. JMV was supported by the Spanish Office for
Science FIS2009-13483-C02-02, Fundaci\'on S\'eneca Regi\'on de
Murcia 11920/PI/09 and the UPCT "Programa de Movilidad". We thank
the University College of London (UK) and the International
Institute of Physics in Natal (Brazil) for their hospitality at
several stages of this project.
\end{acknowledgments}

\bibliography{biblio}

\end{document}